\documentclass{article}
\usepackage{amsmath, amssymb, amsfonts}
\title{Comment on ''The entropy of a hole in spacetime''} 
\author{Hristu Culetu, \\Ovidius University, Dept.of Physics and Electronics, \\B-dul Mamaia 124, 900527 Constanta, Romania, \\e-mail : hculetu@yahoo.com}

\begin{document}
\numberwithin{equation}{section}
\pagenumbering{arabic}
\maketitle
\newcommand{\fv}{\boldsymbol{f}}
\newcommand{\tv}{\boldsymbol{t}}
\newcommand{\gv}{\boldsymbol{g}}
\newcommand{\OV}{\boldsymbol{O}}
\newcommand{\wv}{\boldsymbol{w}}
\newcommand{\WV}{\boldsymbol{W}}
\newcommand{\NV}{\boldsymbol{N}}
\newcommand{\hv}{\boldsymbol{h}}
\newcommand{\yv}{\boldsymbol{y}}
\newcommand{\RE}{\textrm{Re}}
\newcommand{\IM}{\textrm{Im}}
\newcommand{\rot}{\textrm{rot}}
\newcommand{\dv}{\boldsymbol{d}}
\newcommand{\grad}{\textrm{grad}}
\newcommand{\Tr}{\textrm{Tr}}
\newcommand{\ua}{\uparrow}
\newcommand{\da}{\downarrow}
\newcommand{\ct}{\textrm{const}}
\newcommand{\xv}{\boldsymbol{x}}
\newcommand{\mv}{\boldsymbol{m}}
\newcommand{\rv}{\boldsymbol{r}}
\newcommand{\kv}{\boldsymbol{k}}
\newcommand{\VE}{\boldsymbol{V}}
\newcommand{\sv}{\boldsymbol{s}}
\newcommand{\RV}{\boldsymbol{R}}
\newcommand{\pv}{\boldsymbol{p}}
\newcommand{\PV}{\boldsymbol{P}}
\newcommand{\EV}{\boldsymbol{E}}
\newcommand{\DV}{\boldsymbol{D}}
\newcommand{\BV}{\boldsymbol{B}}
\newcommand{\HV}{\boldsymbol{H}}
\newcommand{\MV}{\boldsymbol{M}}
\newcommand{\be}{\begin{equation}}
\newcommand{\ee}{\end{equation}}
\newcommand{\ba}{\begin{eqnarray}}
\newcommand{\ea}{\end{eqnarray}}
\newcommand{\bq}{\begin{eqnarray*}}
\newcommand{\eq}{\end{eqnarray*}}
\newcommand{\pa}{\partial}
\newcommand{\f}{\frac}
\newcommand{\FV}{\boldsymbol{F}}
\newcommand{\ve}{\boldsymbol{v}}
\newcommand{\AV}{\boldsymbol{A}}
\newcommand{\jv}{\boldsymbol{j}}
\newcommand{\LV}{\boldsymbol{L}}
\newcommand{\SV}{\boldsymbol{S}}
\newcommand{\av}{\boldsymbol{a}}
\newcommand{\qv}{\boldsymbol{q}}
\newcommand{\QV}{\boldsymbol{Q}}
\newcommand{\ev}{\boldsymbol{e}}
\newcommand{\uv}{\boldsymbol{u}}
\newcommand{\KV}{\boldsymbol{K}}
\newcommand{\ro}{\boldsymbol{\rho}}
\newcommand{\si}{\boldsymbol{\sigma}}
\newcommand{\thv}{\boldsymbol{\theta}}
\newcommand{\bv}{\boldsymbol{b}}
\newcommand{\JV}{\boldsymbol{J}}
\newcommand{\nv}{\boldsymbol{n}}
\newcommand{\lv}{\boldsymbol{l}}
\newcommand{\om}{\boldsymbol{\omega}}
\newcommand{\Om}{\boldsymbol{\Omega}}
\newcommand{\Piv}{\boldsymbol{\Pi}}
\newcommand{\UV}{\boldsymbol{U}}
\newcommand{\iv}{\boldsymbol{i}}
\newcommand{\nuv}{\boldsymbol{\nu}}
\newcommand{\muv}{\boldsymbol{\mu}}
\newcommand{\lm}{\boldsymbol{\lambda}}
\newcommand{\Lm}{\boldsymbol{\Lambda}}
\newcommand{\opsi}{\overline{\psi}}
\renewcommand{\tan}{\textrm{tg}}
\renewcommand{\cot}{\textrm{ctg}}
\renewcommand{\sinh}{\textrm{sh}}
\renewcommand{\cosh}{\textrm{ch}}
\renewcommand{\tanh}{\textrm{th}}
\renewcommand{\coth}{\textrm{cth}}

\begin{abstract}
Balasubramanian, Czech, Chowdhury and de Boer \cite{BCCdB} studied a ''spherical Rindler space'' and found that accelerating observers are causally disconnected from a spherical region located at the origin of Minkowski space. We show that there is no any horizon of size $R_{0}$ (which is related to the chosen initial conditions) and the event horizon at $r = 0$ is obtained only from the two-dimensional ($r, t$) static restriction. Their near-horizon geometry cancels the time dependence of the original metric and therefore the approximation used is doubtful. \\
\end{abstract}

 In their paper \cite{BCCdB} Balasubramanian et al. (BCCdB) proposed a model for the gavitational entropy of a time-dependent spherically symmetric generalization of ordinary Rindler space. Their entropy is a quarter of the area of the spherical acceleration horizon, as for a black hole. According to the authors of \cite{BCCdB}, the accelerating observers are causally disconnected from a spherical region (a ''hole'') of radius $R_{0}$, located at the origin of Minkowski space. The horizon is spherically symmetric and its size equals $R_{0}$.
 
 Let us notice that BCCdB do not specify any definite acceleration in their equations. Perhaps they have taken it to be unity but that had to be mentioned in some way. For example, the r.h.s. of the 2nd Eq. (2.5) is dimensionless but its l.h.s. is a distance. BCCdB have introduced in their Eq. (5.2) two constants $a$ and $b$ (with $a^{2} = b^{2}$) but they related their meaning to a rescaling of the $\tau$ coordinate. In our view, the best way to assure the same units is to put $gt$ instead of $t$, with $g$ an acceleration. In that case, on their Fig.1 one should consider
  \begin{equation}
  R - R_{0} = r cosh~gt,~~~~T = rsinh~gt
\label{1}
\end{equation}
The above choice is, actually, well-known (see, for instance, \cite{SGP, MP, RL, JWL}). Therefore, the metric (2.6) should be written as
  \begin{equation}
    ds^{2} = -g^{2}r^{2} dt^{2} + dr^{2} + (R_{0} + rcosh~gt)^{2} d \Omega^{2},
\label{2}
\end{equation}
(we restricted ourselves to four dimensions). 

The authors of \cite{BCCdB} stated above Eq. (2.4) that ''...a family of observers accelerating away from a common center, who are causally disconnected from a spherical region of radius $R_{0}$''. However, $R_{0}$ is a value of $R$ measured by a Minkowski observer \footnote{Therefore, the authors' statement ''For spherical Rindler space the horizon is a sphere $R = R_{0}$ in Minkowski space'' (p. 10) seems to contain a misinterpretation: one cannot express a horizon in Rindler space using coordinates from another reference system. For example, we may not say that, for the time dependent Kruskal spacetime, the horizon is the sphere $r = 2m$ in Schwarzschild space. A similar argument works for the time dependent and static de Sitter spacetimes.}. From (0.1) one finds that $R = R_{0}$ corresponds to $r = 0$, which is the only horizon of the accelerated observer. We also notice that the origin of the Minkowski space ($R = 0$) does not coincide with the origin of the Rindler space ($r = 0$). Since $R \geq R_{0}$ always, the topology of the Minkowski space is no longer trivial. But this change emerges because of a mathematical artifact (a hole is assumed to exist) and is not a consequence of some physical process, as is the case of the Rindler observer, to whom the topological change \cite{CD} arises due to a physical phenomenon: the horizon generation by acceleration. 

The fact that the appearance of $R_{0}$ in Eq. (2.6) is a coordinate effect could be justified by means of the following comparison. If the authors of Ref. 1 started with Minkowski metric written in Cartesian coordinates (as Fursaev et al. did in \cite{FPS}) and not in spherical coordinates (2.4), the constant $R_{0}$ would not emerge in Eq. (2.6). Indeed, if we make the Fursaev et al. coordinate transformation (2.11) on their Minkowski metric (2.1), one obtains \footnote{We observe that the constant $a$ will not appear in (2.12) because it cancels by differentiating $x,~y,~z$ from (2.11), as could be verified by direct calculation.} (in their notations) 
  \begin{equation}
     ds^{2} = -r^{2} d\tau^{2} + dr^{2} + r^{2}cosh^{2}\tau d \Omega^{2}. 
 \label{3}
 \end{equation}
 But the role played by their constant $a$ is equivalent to $R_{0}$ from \cite{BCCdB} and so we see how $R_{0}$ is rooted from a coordinate effect.

  For the accelerating observer the horizon is located at $r = 0$ (if we confine to radial motion), namely $R = R_{0} \pm T$ (the light cones) for the Minkowski observer. In our opinion, $R_{0}$ plays no role for the Rindler observer; it represents only an initial condition for the inertial observer who may not see any horizon. Moreover, how did BCCdB find that their geometry (2.6) possess an event horizon of size $ R_{0}$? The geometry is time-dependent and, therefore, there is no a timelike Killing vector to vanish at the horizon, as BCCdB also noticed at the end of Sec. 2. Fursaev et al. gave a prescription to generalise the gravitational entropy for surfaces which are not Killing horizons and the metric is time dependent. Nevertheless, their model does not require any boundaries (as they noticed at the bottom of p.21) and so the entropy emerges as the local property of the surface. In other words, the entropy looses its holographic character. Therefore, we think one is more appropriate here to study the apparent horizon, that is obtained from \cite{HKS, HC1}
  \begin{equation}
 g^{ab} \nabla_{a}\rho ~\nabla_{b}\rho = 0,
 \label{4}
 \end{equation}
with $\rho(r,t) = R_{0} + rcosh~gt$ and $g_{ab}$ are the metric coefficients of their Eq. (2.6). One obtains $g^{ab} \nabla_{a}\rho ~\nabla_{b}\rho = 1$ and there is no any apparent horizon.

For a static Rindler space, it is a known fact that the horizon entropy is actually infinite \cite{RL} because the horizon area is divergent. However, a finite entropy per unit area may be defined. It is worth noting that a finite Rindler horizon entropy could be obtained in terms of the proper acceleration $g$ of the Rindler observer \cite{KM, HC2}, giving $S = \pi/4g^{2}$. 

At the p.8, BCCdB remarked that the only free parameter at their disposal is $R_{0}$, which seems not to be related to a temperature. As we noticed before, there are in fact two parameters entering the equations: $R_{0}$ and the acceleration $g$. In our view, it is $g$, not $R_{0}$, directly related to the horizon temperature, through the Unruh effect. This is also a well-known result, even though the spherical Rindler space is nonstationary. On the other hand, we suppose the near horizon approximation is misused  when one passes from their Eq. (2.6) to Eq. (4.1). The authors of \cite{BCCdB} neglected $rcosht$ w.r.t. $R_{0}$, for $r \rightarrow 0$. Nevertheless, $rcosht$ is exponentially increasing when time advances and the approximation is debatable.

The authors' comparison with the time-dependent de Sitter space is very appropriate: the metric (2.6) and the time-dependent de Sitter metric do not carry any entropy. That takes place, in our view, because of the lack of any horizon for both of them. Let us also observe that the near horizon geometry (4.1) is no longer flat \cite{HC4} due to the approximation used: there is a nonzero scalar curvature $R^{a}_{~a} = 2/R_{0}^{2}$. Moreover, how could the near horizon metric (4.1) become static for any time if the original geometry (2.6) is non-stationary? 

We remark in passage that the metric (2.6) (with $R_{0} = 0$) could be directly obtained from the Minkowski metric in Cartesian coordinates \cite{HC3, FPS}
 \begin{equation}
 X = r cosh gt sin\theta cos \phi ,~~Y = r cosh gt sin \theta sin \phi,~~Z = r cosh gt cos\theta, ~~T = r sinh gt
 \label{5}
 \end{equation}
 Since the nonstatic BBCdB metric (2.6) (we refer to its 4-dimensional restriction) may not have event or apparent horizons, we could slightly modify it to come out an apparent horizon. The idea is to choose a conformal version of it, namely the geometry \cite{HC4, HC5, HC6}
  \begin{equation}
 ds^{2} = \left(1 + \frac{b^{2}}{r^{2}}\right)^{2} \left(-g^{2}r^{2} dt^{2} + dr^{2} + r^{2} cosh^{2}gt d \Omega^{2}\right)
 \label{6}
 \end{equation}
where $b$ is a positive constant of the order of the Planck length. Because in our view $R_{0}$ does not play an important role in the spherical Rindler space, we have taken above $R_{0} = 0$.

The areal radius of the previous geometry is given by 
  \begin{equation}
 \rho (r,t) = \left(1 + \frac{b^{2}}{r^{2}}\right) ~rcosh~gt
 \label{7}
 \end{equation}
Therefore, our Eq. (0.4) leads to 
  \begin{equation}
   - \left(1+\frac{b^{2}}{r^{2}}\right)^{2} sinh^{2}gt + \left(1 - \frac{b^{2}}{r^{2}}\right)^{2}cosh^{2}gt = 0
 \label{8}
 \end{equation}
The apparent horizons appear as
 \begin{equation}
 r_{+} (t) = b~ e^{gt} ,~~~~r_{-} (t) = b ~e^{-gt} ,
 \label{9}
 \end{equation}
In the 1st paper from \cite{HC6} we reached the results (0.9) starting from the condition that the scalar expansions are vanishing \cite{HV}: $\Theta_{\pm} = 0$ on the apparent horizons. It is interesting noting that the two horizons represent null geodesics (expanding/contracting wormhole throat \cite{HC6}). The two null curves (0.9) correspond to those viewed by a spherical distribution of uniformly accelerating observers in Minkowski space, with acceleration $g$.

It is worth noting that Feldman \cite{MF} recently conjectured that inertial motion is characterized by geodesics about ''inertial center points'' in the radial Rindler chart. The geometry (0.2) (with $R_{0} = 0$) plays a central role in his ingenious model.


\begin{thebibliography} {17} 

\bibitem{BCCdB}
V. Balasubramanian, B. Czech, B. D. Chowdhury and J. de Boer, JHEP 1310 (2013) 220 (arXiv: 1305.0856 [gr-qc]).
\bibitem{SGP}
S. Singh, C. Gauguly and T. Padmanabhan, Phys. Rev. D87, 104004 (2013) (arXiv: 1302.7177 [hep-th]).
\bibitem{MP}
B. R. Majhi and T. Padmanabhan, arXiv: 1302.1206 [hep-th].
\bibitem{RL}
R. Laflamme, Phys. Lett. B196, 499 (1987).
\bibitem{JWL}
J.-W. Lee, arXiv: 1011.1657 [hep-th].
\bibitem{CD}
S. M. Christensen and M. J. Duff, Nucl. Phys. B146, 11 (1978).
\bibitem{FPS}
D. V. Fursaev, A. Patrushev and S. N. Solodukhin, arXiv: 1306.4000 [hep-th].
\bibitem{HKS}
S. Hellerman, N. Kaloper and L. Susskind, JHEP 0106: 003 (2001) (ArXiv: 0104180 [hep-th]). 
\bibitem{HC1}
H. Culetu, Class. Quantum Grav. 29, 235021 (2012).
\bibitem{KM}
R. Kerner and R. B. Mann, Phys. Rev. D73, 104010 (2006).
\bibitem{HC2}
H. Culetu, Int. J. Mod. Phys. D15, 2177 (2006); Phys. Lett. B703, 641 (2011).
\bibitem{HC4}
H. Culetu, arXiv: 0903.3548 [hep-th].
\bibitem{HC3}
H. Culetu, AIP Conf. Proc. 1115, 129 (2009) (arXiv: 0804.3754 hep-th]).
\bibitem{HC5}
H. Culetu, Gen.Relat. Grav. 26, 283 (1994).
\bibitem{HC6}
H. Culetu, J. Korean Phys. Soc. 57, 419 (2010); J. Korean Phys. Soc. 57, 17 (2010).
\bibitem{HV}
D. Hochberg and M. Visser, Phys. Rev. D58, 044021 (1998) (ArXiv : gr-qc/9802046).
\bibitem{MF}
M. R. Feldman, arXiv: 1312.1182 [physics.gen-ph].


\end{thebibliography}
\end{document}